\documentclass{elsart}

\usepackage{harvard}

\usepackage{graphicx}

\usepackage{amssymb}

\def\url#1{{\ttfamily\def\/{/\discretionary{}{}{}}#1}}

\begin{document}

\begin{frontmatter}
\title{Electron Population Aging Models for Wide-Angle Tails}


\author[Young]{A. Young\thanksref{ay}}, 
\author[Young]{L. Rudnick},
\author[Katz]{D. M. Katz-Stone},
\author[Donoghue]{A. A. O'Donoghue}

\thanks[ay]{E-mail: ayoung or larry @ast1.spa.umn.edu}

\address[Young]{Department of Astronomy, University of Minnesota, 116 Church St. SE, Minneapolis, MN 55455-0149}
\address[Katz]{US Naval Academy, Annapolis, MD 21402-5026}
\address[Donoghue]{Department of Physics, St. Lawrence University, Canton, NY 13617}
\begin{abstract}

Color-color diagrams have been useful in studying the spectral shapes in radio galaxies. At the workshop we presented color-color diagrams for two wide-angle tails, 1231+674 and 1433+553, and found that the standard aging models do not adequately represent the observed data. Although the JP and KP models can explain some of the observed points in the color-color diagram, they do not account for those found near the power-law line. This difficulty may be attributable to several causes. Spectral tomography has been previously used to discern two separate electron populations in these sources. The combination spectra from two such overlying components can easily resemble a range of power-laws. In addition, any non-uniformity in the magnetic field strength can also create a power-law-like spectrum. We will also discuss the effects that angular resolution has
on the shape of the spectrum.
\end{abstract}
\end{frontmatter}
\section{Introduction}
\label{intro}
We examine the relativistic electron populations in two wide-angle tails 
\linebreak
(WATs): 1231+674 and 1433+553. Previous analyses \cite{Kat99} have identified two apparently distinct populations of relativistic electrons: a flat spectrum 'jet' and a steep spectrum 'sheath'. The overall question for this research is: Can the current aging models account for all of the observed spectra?
\section{Technique}
\label{technique}
We apply two techniques previously developed by Katz-Stone and Rudnick \cite{Kat93,Kat97} to three frequency observations ($\lambda$$\lambda$ 20, 6, and 3.6 cm) taken from the Very Large Array: spectral tomography and color-color diagrams. In this paper, $\alpha$ is defined as I $\propto$ $\nu^{\alpha}$.

Spectral tomography is a technique designed to isolate overlapping structures with unique spectral indices. Two tomography galleries of images are constructed from pairs of frequency maps where (e.g.):
\begin{center}
$I_{\mathrm{t}} = I_{\mathrm{20cm}} - 
(\nu_{\mathrm{20cm}}/\nu_{\mathrm{6cm}})^{\alpha_t} * I_{\mathrm{6cm}}$
\end{center}
Slices are taken on each image in the tomography gallery at the same position. 
The spectral index $(\alpha_{\mathrm{t}})$ of the structure (e.g. the jet) is 
determined when the jet disappears at $I_{\mathrm{t}}$ = background level. 

Color-color diagrams are plots of spectral index at one pair of frequencies vs. spectral index at another pair of frequencies \cite{Kat93,Rud99}. As such, only three frequencies
are needed to plot the local curvature of the spectrum. The locus
of different power-law spectra would lie along the line where 
$\alpha^{1}_{2} = \alpha^{2}_{3}$.
Plotted on the color-color diagram in Fig. 2 are the WAT data for 1231+674 and the three standard aging
models of a synchrotron spectrum. The continuous injection (CI) model assumes that relativistic electrons are constantly supplied into the system. The Jaffe \& Perola (JP) model scatters the relativistic electrons in order to maintain an isotropic pitch angle distributions. The Kardashev \& Pacholczyk (KP) model maintains a constant pitch angle for the relativistic electrons.
\begin{center}
\begin{figure}
\scalebox{.5}[.5]{\includegraphics[-2cm, 5cm][4cm, 22cm]{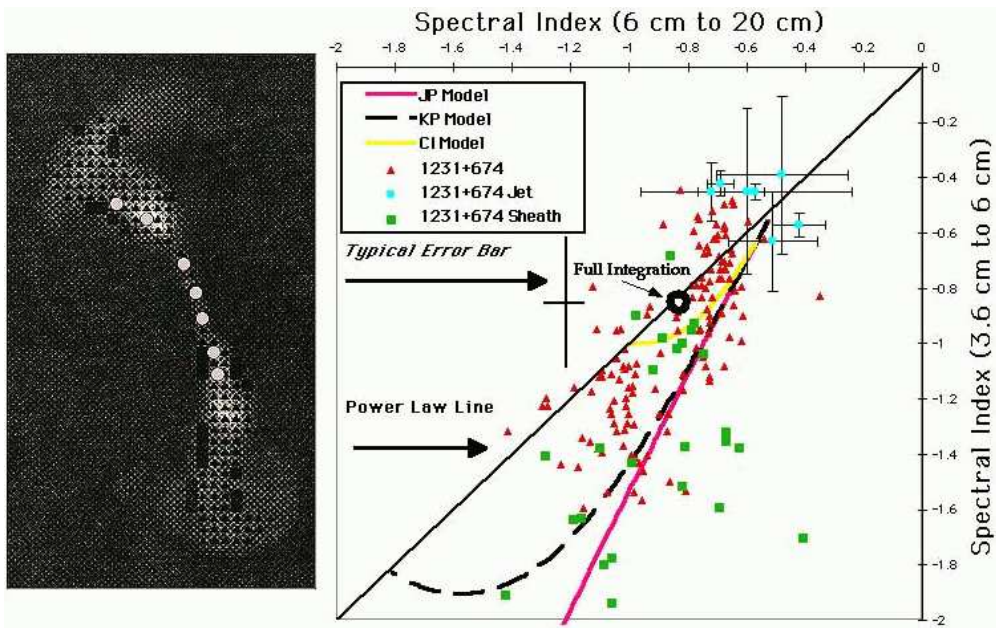}}
\caption{Each point in the image plane ($\lambda$ 20cm) of 1231+674 on the left has a counterpart in the color-color diagram to the right.}
\caption{Color-color diagram of 1231+674 on the right.}
\end{figure}
\end{center}
\label{allcc}
\section{Results and Summary}
\label{Results}
Although the current aging models cannot account for the observed spectra for all positions in 1231+674 or 1433+553 (not shown here), both sources exhibit similar spectral characteristics:
\begin{enumerate}
\item What is the injection spectrum of the jet?

The spectral indices for the jet points (closed circles) in 1231+674 were obtained from the tomography slices. The injection index of the flat spectrum component (jet) follows a power-law with a spectral index of $\alpha$ $\approx$ -0.55 $\pm$ 0.05. The FR I galaxy 3C449 \cite{Kat97} also shows a similar injection spectrum of $\alpha$ $\approx$ -0.53 $\pm$ 0.01. No steepening of the jet's spectrum can be seen up to 56 kpc for 1231+674 and 29 kpc for 1433+553.
\item How do the spectral shapes change with resolution?

The full integrated spectrum (open circle), lies close to a power-law (Figure 2 $\&$ Table 1). However, this integrated spectral index $\neq$ $\alpha_{\mathrm{inject}}$ ($\approx$ -0.55). If we convolve the maps to a lower resolution than shown here in Fig.1, the observed spectra lie close to the power-law line with a broad range of power-laws. This is due to the mixing of spectral components. Again, there is no relation to the apparent injection index ($\alpha$ $\approx$ -0.55) of the jet. If we observe at higher resolution, it may be possible to determine the intrinsic shape of the spectrum for isolated components.
\begin{table*}
\caption{Mean spectral curvature for 1231+674 and (1433+553) at various resolutions.}
\label{tab_ant} 
\begin{center}
\begin{tabular}{c c c c}
\hline 
Source  &  2.7''(2.9'')  &  10.8''(11.6'')  &  Full Integration \\
\hline 
1231+674  & 0.142 $\pm$ 0.23    &  0.10 $\pm$ 0.35  & 0.07 $\pm$ 0.006\\
(1433+553)   & 0.038 $\pm$ 2.96     &  0.025 $\pm$ 0.37  & -0.057 $\pm$ 0.007\\
\hline 
\end{tabular}
\end{center}
\vspace*{.6cm}
\noindent
\end{table*}
\item What is the relationship between the spectral shapes and the local structure of the WATs?

The squares (sheath), obtained from a straight division where $\alpha$ = 
(ln $I_1$ / ln $I_2$) / (ln $\nu_1$ / ln $\nu_2$), were selected based on 
their low surface brightness. This `sheath' population 
(as defined by \citeasnoun{Kat99}) appears to represent a roughly separate 
spectral population in the color-color diagram. In
the color-color diagram, we find that the squares are largely consistent
with a single JP spectrum with an injection index of $\approx$ -0.5 
as measured for the jet. Therefore, the criterion of low surface
brightness appears to be a reliable indicator of an isolated spectral 
population.
The triangles, also obtained from a straight division, represent 
every other 
location on the WATs. However, they are not consistent any of the standard models. They can be simply understood as an apparently broader spectrum caused by a
varying mixture of flat spectrum jet and steep spectrum sheath
components.

Everywhere else in both WATs, the spectral shapes are due to either a mixture of relativisitic electron populations or inhomogeneities in the local magnetic field \cite{Tri93,Tri99}. A single KP or JP model is consistent with many of the spectral regions if we allow for these conditions. In addition, an empirical model presented in \citeasnoun{Kat93} for Cygnus A can also represent much of the spectral shapes seen in the color-color diagrams. For both 1231+674 and 1433+553, the JP model (with $\alpha_{\mathrm{inject}}$ $\approx$ -0.5) serves as an approximate boundary for the maximum curvature seen in the spectrum. 

Alternatively, the observed spectra may be explained by using two homogeneous but distinct electron populations. In this scenario, the jet and sheath begins with different injection indices. Both 1231+674 and 1433+553 would have $\alpha_{\mathrm{jet}}$ $\approx$ -0.55 while $\alpha_{\mathrm{sheath}}$ $\approx$ -1.0 and -0.85 respectively. Future work will include further discussions about these population alternatives and the implications our results may have on current aging analyses.
\end{enumerate}

This work is supported by NSF grant AST 96-16964 to the U. of Minnesota.

\end{document}